\documentclass[10pt,conference,letterpaper]{IEEEtran}
\usepackage{balance}

\usepackage{graphicx} %sead
\usepackage{graphics} %kemal - za resize tabele
\usepackage[english]{babel}
\usepackage[utf8]{inputenc}
\usepackage{mathtools}
\usepackage{tikz}
\usepackage{amsmath,amssymb,amsfonts}
\usepackage{pdfpages}

\usepackage{textcomp}
\usepackage{xcolor}
\def\BibTeX{{\rm B\kern-.05em{\sc i\kern-.025em b}\kern-.08em
    T\kern-.1667em\lower.7ex\hbox{E}\kern-.125emX}}

\usepackage{afterpage}
\usepackage{bm}
\usepackage{color}
\usepackage[font=sc]{caption}

\usepackage{listing}
\usepackage{listings}
\usepackage{fancyhdr}
\usepackage{fancyvrb} 
\usepackage{float}
\usepackage{framed}
\usepackage{enumerate}
\usepackage[shortlabels]{enumitem}
\usepackage{multirow}
\usepackage{hhline}
\usepackage{array}
\usepackage{epstopdf}
\usepackage{pgfplots}
\usepackage{subfigure}
\usepackage{svg}  %kemica - za svg

\usepackage{dblfloatfix}    % To enable figures at the bottom of page

\usepackage{multicol}  % kemica - multiple columns (logy)

\usetikzlibrary{matrix,chains,positioning, decorations.pathreplacing,arrows,plotmarks}
\usetikzlibrary{shapes}
\usetikzlibrary{calc} % for manimulation of coordinates
\pgfplotsset{compat=newest}

\usetikzlibrary{shapes.geometric, arrows}
\tikzstyle{db} = [rectangle, minimum width=0.4\columnwidth, minimum height=0.2\columnwidth, text centered, draw=black, fill=blue!20]
\tikzstyle{file} = [rectangle, rounded corners, minimum width=0.3\columnwidth, minimum height=0.1\columnwidth, text centered,
draw=black, fill=green!20]
\tikzstyle{dir} = [circle, minimum size=0.2\columnwidth, fill=red!20, draw=black]
\tikzstyle{arrow} = [thick, ->, >=stealth]
\tikzstyle{user} = [rectangle, minimum width=0.2\columnwidth, minimum height=0.05\columnwidth, draw=black, fill=black!20]

\usepackage[ruled, vlined]{algorithm2e}

\IEEEoverridecommandlockouts 
\usepackage[OT4,T1]{fontenc}
\usepackage{url}
\usepackage{epstopdf}

\usepackage{cite} % za multicite

\newcommand\MyBox[2]{
  \fbox{\lower0.75cm
    \vbox to 1.7cm{\vfil
      \hbox to 1.7cm{\hfil\parbox{1.4cm}{#1\\#2}\hfil}
      \vfil}%
  }%
}

\title{Knowledge Distillation for Real-Time Classification of Early Media in Voice Communications}
\author{
\IEEEauthorblockN{Kemal Altwlkany $^{1,2}$, Hadžem Hadžić $^{2}$, Amar Kurić $^{2}$, Emanuel Lacic $^{3}$
}
\IEEEauthorblockA{
$^1$Faculty of Science, University of Sarajevo, Bosnia and Herzegovina\\
$^2$Infobip, Sarajevo, Bosnia and Herzegovina\\
$^3$Infobip, Zagreb, Croatia\\
  \{kemal.altwlkany,hadzem.hadzic,amar.kuric,emanuel.lacic\}@infobip.com
}
}

\begin{document}
\maketitle

\begin{abstract}
This paper investigates the industrial setting of real-time classification of early media exchanged during the initialization phase of voice calls. We explore the application of state-of-the-art audio tagging models and highlight some limitations when applied to the classification of early media. While most existing approaches leverage convolutional neural networks, we propose a novel approach for low-resource requirements based on gradient-boosted trees. Our approach not only demonstrates a substantial improvement in runtime performance, but also exhibits a comparable accuracy. We show that leveraging knowledge distillation and class aggregation techniques to train a simpler and smaller model accelerates the classification of early media in voice calls. We provide a detailed analysis of the results on a proprietary and publicly available dataset, regarding accuracy and runtime performance. We additionally report a case study of the achieved performance improvements at a regional data center in India. 
\end{abstract}

\providecommand{\keywords}[1]
{
  \textbf{\textit{Keywords---}} #1
}


\section{Introduction}
\label{sec:intro}

%Standard Voice over IP (VoIP) calls \cite{goode2002voice} are commonly initiated using the Session Initiation Protocol (SIP) \cite{rfc3261}. Similar to the more popular Hypertext Transfer Protocol (HTTP) \cite{rfc2616}, SIP has well defined responses which are identified by a three digit number \cite{rfc3261}. But, when looking at real-world settings, not all telecommunication service providers (TSPs) properly manage SIP responses. Regardless of the actual error at hand, some TSPs continuously propagate the same response and status code through the network.
% This poses an issue for any application logic within a communication platform that depends on VoIP-related services as, e.g., it is unclear if redialing the call should happen and if yes, how many additional attempts should be done.
Standard Voice over IP (VoIP) calls \cite{goode2002voice} are initiated using the Session Initiation Protocol (SIP) \cite{rfc3261}. Similar to the Hypertext Transfer Protocol (HTTP) \cite{rfc2616}, SIP has well defined responses identified by three digit numbers. When looking at real-world settings, however, not all telecommunication service providers (TSPs) properly manage SIP responses. Regardless of the actual error, some TSPs always propagate the same response and status code through the network. 
This poses an issue for any application logic within a communication platform that depends on VoIP-related services.
%Aside from the obvious drawbacks of not following protocol standards, this poses an issue to developers whose application logic depends on knowing the status code of the response as error handling is a crucial part of any communication platform. 

%A key metric in assessing the overall network quality here is the Answer Seizure Ratio (ASR) \cite{ITUrecE425}. ASR is characterized as the ratio between successful call connections and total call attempts. Enhancing ASR involves strategies like redialing failed connections or using alternate gateways to establish the call, especially during network congestion or poor signal scenarios. However, excessive redialing is impractical, as it can overload the network and lead to user suspension for spam-like behavior. All of this highlights the need to address the lack of suitable SIP response codes in such situations.
A key metric in assessing the overall platform quality here is the Answer Seizure Ratio (ASR) \cite{ITUrecE425}, defined as the ratio between successful call connections and total call attempts. Enhancing ASR involves strategies like redialing failed connections or using alternate gateways to establish the call, especially during network congestion or poor signal scenarios. However, excessive redialing can overload the network and lead to user suspension for spam-like behavior. All of this highlights the need to address the lack of suitable SIP response codes in such situations.
An interesting observation, however, is that TSPs which lack the proper SIP response codes for failed calls often provide some form of feedback through early media.
Early media is defined in the RFC standard as media that is exchanged before a particular session is accepted by the called party \cite{rfc3960}. In many cases, it can be possible to map an announcement received as early media onto an appropriate SIP code, as the announcement might indicate that the callee is busy, not available, out of network coverage, or similar. 
%Moreover, the exchange of early media is commonly facilitated through the real-time transport protocol (RTP) \cite{rfc3960, rfc3550}.

%As such, the focus of this paper is in analyzing the received early media stream and determining whether it can be matched with an appropriate SIP code. This process, although very effective in terms of improving the ASR, can easily become computationally expensive when analyzing every received early media stream, especially as not all received media contain an announcement that can be mapped onto a SIP code. 
%Aside from announcements or IVRs, a very common utilization of early media is for it to send ringing \cite{rfc3960}. 
%For example, early media can sometimes contain music files, or comfort noise \cite{rfc3960}. 
%Therefore, not every early media stream must be analyzed in order to map a potential announcement onto a SIP code. 
%As only announcements should be analyzed and not other IVR types, the task at hand is to distinguish between a coarser group of audio types such as: ringing, music, noise and any type of speech. 
The process of mapping an early media announcement to a corresponding SIP code is very effective in terms of improving the ASR. However, analyzing every early media stream is computationally expensive, as well as unnecessary - not all early media streams contain an announcement that can be mapped onto a SIP code. 
%Aside from announcements or IVRs, a very common utilization of early media is for it to send ringing \cite{rfc3960}. 
For example, early media sometimes carries music files or comfort noise \cite{rfc3960}. 
%Therefore, not every early media stream must be analyzed in order to map a potential announcement onto a SIP code. 
Since only announcements should be analyzed, this paper focuses on the task of classifying audio into the following coarse classes: ringing, music, noise and human speech. This makes it possible to perform further analysis only when necessary, i.e. when human speech is detected.

\section{Related Work}
\label{sec:relwork}
Regarding the task of audio classification, Bugatti et al. \cite{bugatti2002audio} compared a statistical approach to a neural network for determining whether audio carries speech or music. The statistical approach was based on the zero-crossing rate and Bayesian classification, while the neural network approach relied on a simple MLP with 8 input features \cite{bugatti2002audio}. Although the MLP approach offers better performance, it comes at the expense of increased computational complexity. This is highly aligned with our motivations for selecting a GBT model.
%\ka{What was the outcome? What did the author show which approach was better and maybe why?} \textbf{EDIT: The authors concluded that the statistical approach is computationally more efficient due to the simplicity of the model, but the MLP approach offers better performance at the expense of a limited growth in the computational complexity \cite{bugatti2002audio}. This is highly aligned with our motivations for selecting a gradient-boosted trees model.}
Similarly, Lavner and Ruinskiy \cite{lavner2009decision} presented a decision-tree-based algorithm that classifies audio into music and speech.
The authors of \cite{choi2016automatic} showed how to train a deep convolutional neural network in order to tag audio into 50 most common tags from the Million Song Dataset \cite{bertin2011million}.
%, which is similar to gradient-boosted trees used in our work. 
%\textbf{EDIT: Choi et al. \cite{choi2016automatic}  trained a deep convolutional neural network in order to tag audio into 50 most common tags from the Million Song Dataset \cite{bertin2011million}.}
%\ka{How is Choi a broader classification problem? Because of the number of classes in the Million Song dataset?}
%In a similar manner, \cite{cakir2015polyphonic} uses deep neural networks for sound event detection. 
When looking at state-of-the-art neural network models, YAMNet \cite{tensorflowmodelgarden2020} and PANNs \cite{kong2020panns} need to be noted.  
YAMNet was trained on AudioSet-YouTube corpus \cite{audioset}, can classify audio into 521 distinct classes 
%and contains about 3.75 million parameters 
and operates on segments of $0.96$ seconds of length, with a hop length of $0.48$ seconds \cite{mohammed2023radio}. 
% YAMNet was trained on the AudioSet-YouTube corpus \cite{tensorflowmodelgarden2020,audioset}. 
%Kong et al. \cite{kong2020panns} presented PANNs, a collection of neural networks for audio tagging, also trained on the AudioSet dataset \cite{audioset}. They refer to these models as pretrained audio neural networks (PANNs) and obtain a state-of-the-art mean average precision (mAP) of 0.439 on AudioSet with their best model, thereby outperforming the previous best system \cite{kong2020panns}.
 PANNs (i.e., pretrained audio neural networks) is a collection of neural networks for audio tagging that are also trained on the AudioSet dataset. The best model reports a state-of-the-art mean average precision (mAP) of $0.439$ on 527 different classes.

\section{Methodology}

In what follows we describe the process of training a simple gradient-boosted decision tree using two transfer learning concepts: class aggregation and knowledge distillation.

\subsection{Knowledge Transfer}
As previously mentioned, there is a body of research regarding the task of audio classification. A widely used dataset is AudioSet \cite{audioset}, which is composed of over $5,500$ hours of audio labeled with $527$ different classes. These classes cover a wide variety of sounds, some being very specific. As the classification of these sounds can be a hard problem to tackle, state-of-the-art solutions mostly employ deep neural networks.
However, their size restricts their applicability to early media classification, as a fast and resource-efficient classifier is expected to handle a large amount of parallel calls in real-time.

\vspace{2mm} \noindent 
\textbf{Class aggregation.} 
In order to be able to transfer the knowledge on audio classification from methods like \cite{kong2020panns} and make it suitable to be applied in an industrial setting for early media classification, we propose to map each of the original $527$ classes to one of 4 classes of interest: announcement, ringback, music or silence. This mapping significantly reduces the complexity of the problem, and therefore allows a simpler and more cost-efficient model to achieve high performance.% During the creation of our dataset we have  omitted all audio samples that we weren't able to unambiguously map onto the reduced set of classes.
%  \textcolor{red}{In cases when applying. for a newly seen audio file, if a state-of-the-art neural model would predict an AudioSet class that does not correspond to any of the four labels, we would ignore this audio for training a model.

%\vspace{2mm} \noindent 
%\textbf{Real-world requirements.} 
%Based on the industrial setting to which our model will be deployed, it is only expected to handle PCMA and PCMU encoded audio \cite{ITUrecG711}. Thus, the sampling rate of recordings used to form the dataset for training a model would need to be $8$ kHz \cite{ITUrecG711}.% In cases when a different sampling rate is needed, either re-sampling prior to the inference could be considered or a new model would need to be trained in order to fit the required sampling rate.

\vspace{2mm} \noindent 
\textbf{Knowledge distillation.} 
Our aim is to transfer the knowledge from the more complex neural models like the one described in \cite{kong2020panns}. To be more specific, in this paper we utilize CNN14 which was the best performing model (a $14$-layer convolutional neural network). 
% to classify the collected early media recordings. 
For an input audio recording with a duration of $n$ seconds, this neural network returns $100 \times n$ class predictions, one for each $10$-millisecond window. However, inconsistencies in the returned classes can commonly occur. For example, the network may infer one second of speech, followed by $20$ milliseconds of music, followed again by speech. In order to combat such issues, a class-smoothing procedure was employed where each class was compared with the most common class in its neighborhood.

Denote by $L_i$ the class (one of the aggregated four) assigned by the CNN14 model to frame $i$ of an audio recording. For each class $L_i$, a Hann window of size $301$ is centered on $L_i$, so that it spans classes from $L_{i-150}$ to $L_{i+150}$. Classes in this range are assigned weights corresponding to the coefficients of the window. In other words, class $L_{i \pm k}$ is assigned weight: 

\begin{equation}
    W_k = 0.5-0.5\cos\left(\frac{2\pi(150+k)}{300}\right).
\end{equation}

Class $L_i$ is then replaced by the class in its neighborhood which has the largest cumulative weight:

\begin{equation}
L^{new}_i = \arg\max_{l} \sum_{-150 \le k \le 150} I(i+k, l) \cdot W_k,
\end{equation}

where $L^{new}_i$ represents the newly assigned class, and $I(t, l)$ is $1$ if $L_t = l$ and $0$ otherwise. These newly assigned classes are then used as ground truth for training a smaller model.

\subsection{Gradient-Boosted Decision Tree Model}
%As mentioned, when applying the above procedure for newly seen early media recordings using CNN14, we end up with classes for each $10$-millisecond window. To train a simpler model, we focus on classifying a single second of audio into announcement, ringback, music or silence. To accommodate for this, whenever a given early media audio contains a consecutive second (i.e. $100$ consecutive classes) belonging to the same class, that second of audio is used to create a datapoint which can be used for training.% This is yet another example of the simplification that was done, enabling reduction of model complexity since we do not require such a high resolution in time.
\noindent \textbf{Time resolution.} As mentioned, CNN14 predicts a class for each $10$-millisecond window of input audio. To train a simpler model, we instead focus on classifying a single second of audio. To accommodate for this, whenever a given early media recording contains a consecutive second (i.e. $100$ consecutive frames) belonging to the same class, that second of audio is used to create a datapoint which can be used for training.% This is yet another example of the simplification that was done, enabling reduction of model complexity since we do not require such a high resolution in time.

\vspace{2mm} \noindent 
\textbf{Cost-effective model.} 
Since cost-efficiency is a concern, using a GPU in production for complex neural architectures is not a viable option. As such, in this paper we employ Gradient-Boosted Decision Tree (GBT) as the machine learning algorithm. We train the model using Microsoft's LightGBM framework \cite{ke2017lightgbm}. 
%The main advantage of this algorithm over using CNN14 directly is a reduction in inference time by an order of magnitude, which we show in more details in the following sections.

\vspace{2mm} \noindent 
\textbf{Training dataset.} We utilize a proprietary dataset of early media recordings, which were collected in the same environment where the audio classification needs to take place. Due to the nature of our data, we are not able to publicly source the dataset used for training. However, we provide the statistics of the used data, as well as report a benchmark of our model when evaluated on the popular and publicly available AudioSet \cite{audioset} dataset in the next section.
Our dataset contains $135,000$ seconds of music, $54,000$ seconds of ringing, $158,000$ seconds of silence and $201,000$ seconds of announcements. We employed a stratified $80$/$20$ split to create a train and test set (i.e. $20\%$ of recordings for each class were incorporated into the test set).% The split was random, although care was taken to prevent segments from the same recording being present in both the train and test sets, as this could impact the accuracy evaluation.

\vspace{2mm} \noindent 
\textbf{Input features.} 
Instead of feeding the model with raw audio samples, the audio is preprocessed in order to get a more representative set of features with lesser dimensionality. There are multiple ways to do this, but a common approach is by utilizing either the power mel-frequency spectrogram or mel-frequency cepstral coefficients \cite{nour2008mel}. When computing the spectrogram, we use a hop-length of $31.25$ milliseconds, which results in $29$ timesteps within one second. The number of mel bands is $24$, meaning the overall input is a matrix of size $29 \times 24$. This results in a total of $696$ input features, regardless of whether the mel-frequency spectrogram is used, or the mel-frequency cepstral coefficients.

\vspace{2mm} \noindent 
\textbf{Hyperparameter and feature analysis.} We experimented with both mentioned input feature types: power mel-frequency spectrogram and mel-frequency cepstral coefficients (MFCC). MFCCs are obtained by applying two additional transformations to the mel-frequency spectrogram, that is, by calculating the log-spectrogram which is then followed by applying the discrete cosine transform (DCT). Applying DCT has the effect of decorrelating the obtained mel bands, which was shown to work better for some machine learning algorithms \cite{nour2008mel}.
% In both cases, a grid search was performed to find the best hyperparameter values. The parameters that were tuned are number of leaves and minimum data per leaf. In total, $30$ different parameter combinations were tested using $5$-fold cross validation for each set of input features. We also employed early stopping to prevent overfitting.

Our experiments resulted in no significant difference in classification accuracy  between the mel-frequency and MFCC inputs, i.e., we have not found a clear benefit in introducing additional operations required to obtain MFCCs.

  %Hence, to apply our approach in real-time, we utilize mel-frequency spectrograms as the form of input for the model. Furthermore, the grid search indicated that the hyperparameters with the lowest error rate were $31$ for number of leaves and $100$ for minimum data in leaf. These values were selected as the hyperparameters  for our final GBT model.

\vspace{2mm} \noindent

\section{Results}

We are interested in several performance metrics of our model. First of all, we wish to compare the accuracy of our model with CNN14 on our proprietary dataset (which most closely resembles the data the model will be encountering in a production environment). This indicates the quality of the transferred knowledge from CNN14 to GBT. To compensate for the dataset used to distill the knowledge not being publicly available, we provide a head-to-head comparison of the model's performance on a small subset of the AudioSet dataset \cite{audioset}. Lastly, we are interested in the runtime performance of our model, thus we compare the inference times of GBT and CNN14.

\begin{table}[t!]
\setlength{\tabcolsep}{12pt}
\Large
\renewcommand{\arraystretch}{1.5} % Adjusts the row height
\center
\caption{Confusion matrix of predicted early media classes when comparing CNN14 and the simple GBT model.}
\resizebox{\columnwidth}{!}{%
\begin{tabular}{ll||c|c|c|c}
\multicolumn{2}{c||}{}&\multicolumn{4}{c}{CNN14}\\

\multicolumn{2}{c||}{}&Music&Ringing&\multicolumn{1}{c|}{Silence}&\multicolumn{1}{c}{Announcement}\\
 \hline \hline
\multirow{4}{*}{GBT}
&  Music & $26,779$ & $1$ & $2$ & \multicolumn{1}{c}{$331$}\\
\cline{2-6}
& Ringing & $6$ & $13,617$ & $0$ & \multicolumn{1}{c}{$10$}\\
\cline{2-6}
& Silence & $59$ & $1$ & $39,486$ & \multicolumn{1}{c}{$164$}\\
\cline{2-6}
& Announcement & $193$ & $106$ & $9$ & \multicolumn{1}{c}{$50,133$}\\
\hline 
\end{tabular}

}

\label{tab::confusion matrix test set}
\end{table}

\subsection{Performance on proprietary dataset}
With respect to the prediction performance after distilling the knowledge from the more complex CNN14 to the simple GBT, we report our results in Table \ref{tab::confusion matrix test set}. Here we show the confusion matrix on the \textit{test set} when comparing the predicted classes between GBT and CNN14. We are able to achieve very little deviation from the performance of CNN14, which in our case serves as the ground truth for the unlabeled early media from our proprietary dataset. Overall, the simple GBT model achieves a prediction performance that overlaps by $99.3\%$ with CNN14 which is analogous to the model's accuracy. The most common mismatch is happening between music and announcement classes, which is expected since certain songs have parts that closely resemble speech (e.g. rap music or acapella). The per-class precision and recall are well above 99\% for all classes, except the recall of music, which equals a slightly lesser value of 98.7\%. This indicates that CNN14's knowledge of classifying audio into 527 classes has successfully been distilled onto GBT for classifying audio into an aggregated set of 4 classes.

\begin{figure}[!t]
    \centering
    \includegraphics[width=0.80\linewidth]{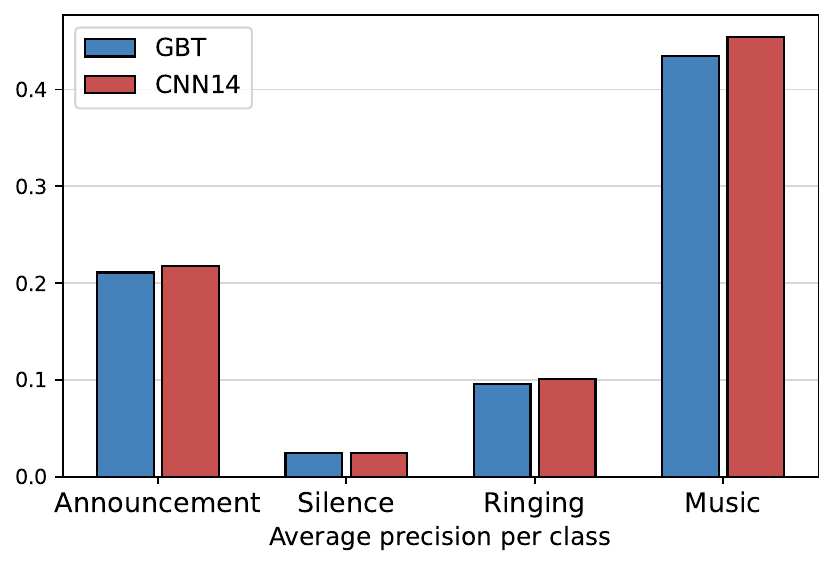}
    \caption{Average precision per class of GBT and CNN14 evaluated on AudioSet \cite{audioset}. The mean average precision (mAP) of GBT is 0.193 and the mAP of CNN14 is 0.2 overall.}
    \label{fig::mean-average-precision}
\end{figure}

\begin{figure*}[!t]
    \centering
    \includegraphics[width=1\linewidth]{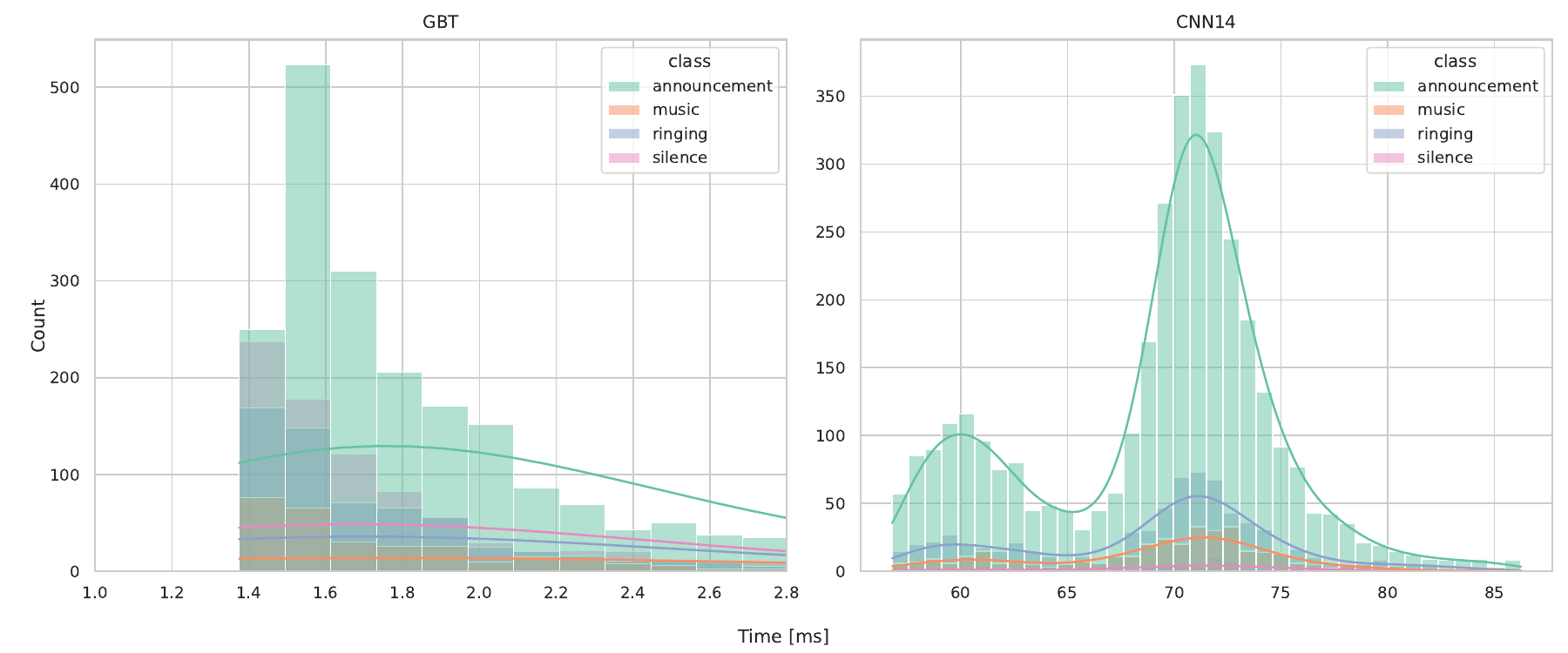}
    \caption{Inference time of simple GBT (left) and CNN14 (right) models, aggregated per class. The GBT model has a significant reduction in inference time by a factor of about 39. The inference time of both models is invariant to the class of early media.}
    \label{fig::inference-time}
\end{figure*}

\subsection{Performance on AudioSet}
Even though the model is specifically designed to classify early media files, we performed an additional experiment on a subset of AudioSet \cite{audioset} data. Using the training subset of AudioSet would not be representative as CNN14 was already trained on it. AudioSet's test subset contains approximately $20,000$ audio files, but we verify the model performance only on files which could be unambiguously mapped onto the aggregated classes: \textit{music}, \textit{ringing}, \textit{silence} or \textit{announcement}, resulting in a total of 550 audio files. The average precision per each class of both the CNN14 and GBT models obtained on this subset of AudioSet are presented in Figure \ref{fig::mean-average-precision}. Both models attain a relatively similar mean average precision (mAP) score, 0.193 for GBT and 0.2 for CNN14.

There are a few interesting conclusions to be drawn from the obtained results.
The performance of the GBT model is very similar to that of the CNN14, but slightly lower per each class, indicating that the knowledge has been transferred successfully. Both models achieved a very poor average precision for classes \textit{silence} and \textit{ringing}, compared to classes \textit{music} and \textit{speech}. This is expected since CNN14 itself struggles with detecting silent frames, as can be confirmed in the appendix of \cite{kong2020panns}, where the per-class average precision is the lowest for class \textit{silence}, and is very low for class \textit{ringing} as well.

\subsection{Runtime performance}
We additionally conducted a small experiment on $10,000$ early media recordings with a mean duration of $5$ seconds (i.e., about $50,000$ segments of 1 second audio) and compared the inference times between CNN14 and the trained GBT model. The results are shown in Figure \ref{fig::inference-time}. Here we deployed both models on a CPU, specifically the Intel Core i7-1165G7. For $1$ second of input audio, the median inference time of CNN14 is about $68$ milliseconds, while the GBT model takes only about $1.74$ milliseconds (both values including preprocessing). This is a relative reduction in inference time by a factor of about $39$.

%It is worth noting that the CNN14 architecture provides a wider distribution of the runtime, whereas the GBT model is more consistent, as also seen in Figure \ref{fig::inference-time}. 
The inference time of both models is invariant with respect to the type of early media (e.g. speech or music). This is an expected behavior of the CNN14 model as the layers of the neural network perform the same operations regardless of the type of input. We wanted to verify the behavior of the GBT model, as for some classes the underlying decision trees of the model could terminate earlier/later. However, this turned out not to be the case.

%Beside the runtime improvement due to a simpler model, another advantage of our proposed approach is not needing to sample the input audio at $32$ kHz, which is a requirement for CNN14. Applying CNN14 directly would therefore require upsampling the input audio, which would incur additional computational costs.

%\input{figs/map}

\section{Case Study}
To illustrate the significance of the acceleration achieved through our proposed method, we also report on concrete online performance results from a single Infobip data center, located in the country of India. Here we analyzed the traffic during a $24$-hour period which utilized the early media classifier. A total of $626,364$ early media files were processed. Given the variability of the duration of early media, the total number or processed segments by the GBT model accumulated to $17,078,525$. By employing our GBT model, we are able to differentiate announcements from other types of early media, thus running the more costly SIP code mapping procedure only when necessary. Given that the percentage of announcements amounted to around 25\%, we managed to reduce the average processing time of $4$ seconds of early media from $77.81$ milliseconds to $25.05$ milliseconds. This is an improvement by a factor of $3.11$, which is significant given the large volume of traffic on a single data center. The improvement has made it possible to serve on average $3$ times more parallel VoIP calls per single CPU core, thereby significantly reducing operational costs.

\section{Conclusion}
%In this paper, we tackled the problem of real-time early media audio classification in voice communications. We explored the application of a simple model when trained on ground truth that is derived using more complex state-of-the-art neural models. Our approach addresses two key aspects: (1) the viability of using simple classifiers in a low-resource setting and, (2) the effectiveness of such a classifier in accurately identifying early media audio types. We showed how to operate efficiently in resource-constrained environments, making a simple model suitable for real-time applications. Furthermore, our results demonstrate that a simple classifier can effectively distinguish between different types of early media audio, thereby enhancing the accuracy of audio classification in voice communication systems. For future work, we aim to further improve the classification accuracy by combining additional simple models within a hybrid setting. Additionally, we plan to investigate the application of our approach in different real-world scenarios and its impact on system performance.
This paper tackled the problem of real-time classification of early media audio in voice communications.
We showed how to train a much simpler gradient-boosted trees model through knowledge distillation from state-of-the-art neural audio tagging models. Our findings confirm that a simple model can operate efficiently in resource-constrained environments, making it suitable for real-time applications.
This methodology effectively balances the need for real-time performance and high accuracy which is usually achieved by more complex neural architectures. 
For future work, we aim to further improve the classification accuracy and lower the required input audio duration by combining additional simple models within a hybrid setup. 
%We are especially interesting in lowering the requirement of the duration of the input audio to be bellow $1$ second.

% add fw based on knowlegde distilation from: TOWARDS DOMAIN GENERALISATION IN ASR WITH ELITIST SAMPLING AND ENSEMBLE KNOWLEDGE DISTILLATION ?

\balance

\bibliographystyle{ieeetr}
%\vspace*{3cm}
\bibliography{references.bib}

\end{document}